\documentclass[twocolumn,showpacs,preprintnumbers,amsmath,amssymb,epsf,superscriptaddress]{revtex4-1}
\usepackage{graphicx,epsfig}% Include figure files
\usepackage{dcolumn}% Align table columns on decimal point
\usepackage{bm}% bold math
\begin{document}

%\preprint{ICN/000-HEP}

\title{Cosmic-Ray Hardenings in the Light of AMS-02}
\author{Yutaka Ohira}
\email{ohira@phys.aoyama.ac.jp}
\affiliation{Department of Physics and Mathematics, Aoyama Gakuin University, 5-10-1 Fuchinobe, Sagamihara 252-5258, Japan}

\author{Norita Kawanaka}
\affiliation{Department of Astronomy, Graduate School of Science, University of Tokyo, 7-3-1, Hongo, Bunkyo-ku, Tokyo, 113-0033, Japan}

\author{Kunihito Ioka}
\affiliation{Theory Center, Institute of Particle and Nuclear Studies, KEK, Tsukuba 305-0801, Japan}
\affiliation{Department of Particle and Nuclear Physics, SOKENDAI (the Graduate University for Advanced Studies), Tsukuba 305-0801, Japan}

%\date{\today} % It is always \today, today,

\begin{abstract}
Recent precise observations of cosmic rays (CRs) by AMS-02 experiment clearly show (1) harder spectra of helium and carbon compared to protons by $\propto R^{0.08}$, and (2) concave breaks in proton and helium spectra at a rigidity $R \sim 300$ GV. 
In particular the helium and carbon spectra are exactly similar, pointing to the same acceleration site.
We examine possible interpretations of these features and identify a chemically enriched region, that is, superbubbles as the most probable origin of Galactic CRs in high rigidity $R>30~{\rm GV}$. 
The similar spectra of CR carbon and helium further suggest that the CRs with $R>30~{\rm GV}$ 
originate from the supernova ejecta in the superbubble core, mixed with comparable or less amount of interstellar medium. We predict similar spectra for heavy nuclei. 
\end{abstract}

\maketitle
%%%%%%%%%%%%%%%%%%%%%%%%%%%%%%%%%%%%%%%
%%%%%%%%%%%%%%%%%%%%%%%%%%%%%%%%%%%%%%%
\section{Introduction}
Supernova remnants (SNRs) are thought to be the origin of Galactic cosmic rays (CRs).
Yet, we have not understood what type of supernova dominantly produces Galactic CRs in each energy, 
e.g., isolated supernovae or supernovae in superbubbles \cite{Drury:2012md}. 
There are many suggestions and interesting discussions in previous studies about CR compositions, that are based on CR data in the rigidity $R<1~{\rm GV}$ \cite{Meyer99}. 
Now, we consider high energy CRs with $R\gg1~{\rm GV}$. 
If the spectral index depends on CR species, the CR composition should depend on the rigidity, $R$.

\begin{figure}
{\centering
{\includegraphics[width=8.6cm]{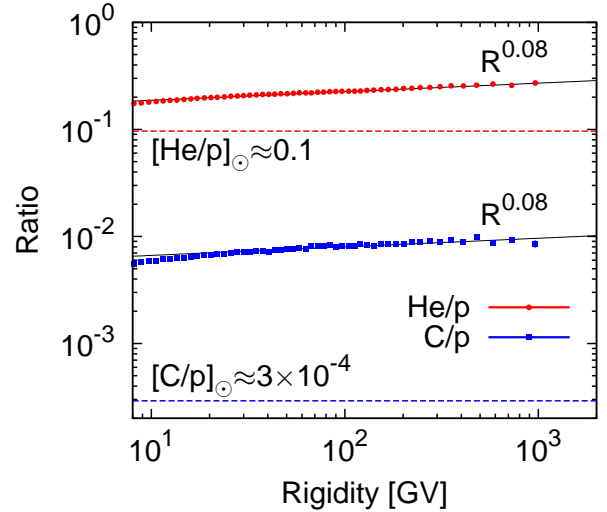}}
\par}
\caption{The helium to proton ratio (He/$p$; {\it filled circle}) and the carbon to proton ratio (C/$p$; {\it filled square}) as a function of rigidity $R$, that is provided by AMS-02 \cite{aguilar15}.
These ratios are not constant but clearly rise as $R^{0.08}$.
In addition, these ratios are appreciably larger than the solar abundance ({\it dashed lines}).
On the other hand, the carbon to helium ratio is constant.}
\label{fig:ratio}
\end{figure}

Recently accurate measurements by AMS-02 experiment \cite{aguilar15} greatly improve previous results by PAMELA \cite{Adriani:2011cu}, CREAM \cite{Ahn:2010gv}, AMS and ATIC-2.
Two kinds of ''CR hardening'' are observed 
by AMS-02:
\begin{enumerate}
\item Spectra of CR helium and carbon are harder than that of protons, and the spectral index of CR carbon is almost the same as that of CR helium (see Figure~\ref{fig:ratio}).

\item The spectra of CR protons and helium have an upturn break at a rigidity $R\sim 300~{\rm GV}$ (see Figure~\ref{fig:spec}). The CR carbon possibly has the same break although the data is not sufficient.
\end{enumerate}

The precision measurements of carbon up to $R\sim1~{\rm TV}$ are newly provided by AMS-02. 
From Figure~\ref{fig:ratio},
we find that the ratios, He/$p$ and C/$p$, are well approximated by the power law as $\propto R^{0.08}$.
Although the spallation and solar modulation reduce helium and carbon at low energy,
they distort the power law by less than $\sim 20\%$ in the energy range of $10$--$10^3$ GV \cite{Vladimirov:2011rn}, where the effect of spallation may be estimated from the boron-to-carbon ratio (B/C).
Remarkably, the CR carbon has the same spectrum as the CR helium, which strongly suggests that both carbon and helium are accelerated at the same place (see Section~\ref{sec:inhomo} for detail). 
Furthermore, the abundances of the CR carbon and helium are larger than the solar abundance \cite{Lodder:2003zy}.
Especially for the CR carbon, the abundance is over 10 times larger.

Before AMS-02, several models were proposed to explain the helium hardening. 
The new AMS-02 data, especially of carbon, can narrow down these models and the CR sources.
In this paper, we discuss which model is better to explain the new data and 
conclude that the superbubble model is most likely (see Table~\ref{tab:score1}).
Therefore, we first briefly show that 
our superbubble model \cite{Ohira:2010eq} can explain the new data of AMS-02 (Section~\ref{sec:2}). 
Then, we discuss which model is better to explain several features of observed CR spectra, i.e., the overabundance of CR carbon (Section~\ref{sec:carbon}), the different spectra between CR protons and CR helium/carbon (Section~\ref{sec:4}, Table~\ref{tab:score1}), and the spectral breaks at $R\sim 300~{\rm GV}$ (Section~\ref{sec:5}).
Section~\ref{sec:6} is devoted to summary and discussion. 

Note that AMS-02 also provides antimatter data.
Here we assume that the antiprotons and positrons are produced by other sources, 
such as nearby SNRs \cite{Kohri:2015mga}. 
If nearby SNRs interact with dense molecular clouds, secondary antiprotons and positrons originated from the SNRs can dominate the Galactic component, while CR protons from them fall below the Galactic component. 

%%%%%%%%%%%%%%%%%%%%%%%%%%%%%%%%%%%%%
%%%%%%%%%%%%%%%%%%%%%%%%%%%%%%%%%%%%%
\section{superbubble model}
\label{sec:2}
\begin{figure}
{\centering
{\includegraphics[width=8.6cm]{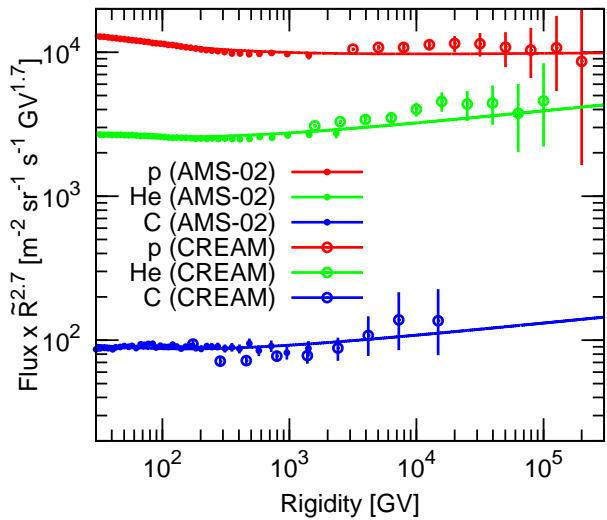}}
\par}
\caption{Spectra of CR protons (red), helium (green), and carbon (blue) fluxes as a function of rigidity. The data are given by AMS-02 \cite{aguilar15} and CREAM \cite{Ahn:2010gv}. These are fitted by theoretical curves based on the superbubble model in Section 2 (sold lines).}
\label{fig:spec}
\end{figure}
\begin{table*}
\caption{Score Sheet for Models of Different Cosmic-ray Spectra}
\label{tab:score1}
\begin{tabular}{lcclc}
\hline \hline
Model  &  Mechanism & Score & Comment & Reference \\
\hline
Propagation & Spallation &  C & Inconsistent with data of B/C & \cite{Blasi:2011fi} \\
\hline
Different sources  & Wind & C & Requires suppression of He (p) acceleration in the ISM (winds) & \cite{Biermann:2010qn}  \\
                             & Reverse shock & B & Requires suppression of He acceleration in the ISM &  \cite{Ptuskin:2012qu} \\
\hline
 Injection   &  Injection & C & Inconsistent with C/He=const. & \cite{Malkov:2012gb} \\
  \hline
 Inhomogeneous environment& Ionization & C & Inconsistent with C/He=const. & \cite{Drury:2010am} \\
                                       & Superbubble & A  & Consistent with observations, and CRs originate from ejecta. &\cite{Ohira:2010eq} \\
\hline \hline
\end{tabular}
\end{table*}
According to the standard theory of Galactic CRs, CRs are accelerated by SNRs.
The CRs escape from SNRs to the interstellar medium (ISM), propagate in our Galaxy,
and finally escape from our Galaxy.
The steady-state spectra of CR protons, helium and carbon are given by 
\begin{equation} 
\frac{dN_{i}}{dR} \propto R^{-s_{{\rm obs},i}} , s_{{\rm obs},i} = s_{{\rm inj},i}  + \gamma_i ~~,
\label{eq:sobs}
\end{equation}
where the subscript $i=p, {\rm He}, {\rm C}$ represents particle species and $\gamma_i$ is the rigidity dependence of the diffusion coefficient in our Galaxy, $D\propto R^{\gamma_i}$. 
$s_{{\rm inj},i}$ is the spectral index of CRs injected into the ISM. 

High energy CRs are thought to be accelerated in SNRs and escape from them earlier than low energy CRs 
because the shock velocity is fast and the magnetic field is strongly amplified in the early phase. 
On the other hand, low energy CRs are mainly accelerated in SNRs at later times because 
the mass swept by the shock of SNRs increases with time and CRs injected early lose their energy by the adiabatic expansion before they escape. 
Therefore, the spectra of CRs inside SNRs are given by 
\begin{equation} 
\frac{dN_i}{dR} \propto t^{\beta_i}R^{-s_i} ~~,
\end{equation}
and the time evolution of the maximum rigidity is given by 
\begin{equation} 
R_{{\rm max},i} \propto t^{-\alpha_i} ~~.
\end{equation}
$\alpha_i$ and $\beta_i$ are parameters to describe the evolution of the maximum rigidity and the number of CRs, respectively, and $s_i$ is the spectral index of CRs inside SNRs.
By taking account of the rigidity dependence of CR escape from sources \cite{ohira10}, 
the index, $s_{{\rm inj},i}$, is given by  
\begin{equation} 
s_{{\rm inj},i} = s_i  + \frac{\beta_i}{\alpha_i} ~~. 
\label{eq:sinj}
\end{equation}
If all CRs originate from the same source, the physics parametrized by $s_i$, $\alpha_i$ and $\gamma_i$ depends only on particle rigidity. 
Then, the inferred rigidity spectrum should be universal. 
However, the number evolution parameter, $\beta_i$, could depend on the particle species 
if the composition distribution is not uniform because $\beta_i$ depends on the 
density distribution around SNRs. 
If so, the steady-state spectra of CRs (Equation~\ref{eq:sobs}) depend on the CR species. 

In addition, taking account of the rigidity dependence of CR escape from sources \cite{ohira10}, 
the spectral index, $s_i$, in Equation (\ref{eq:sinj}) may depend on the rigidity 
because it depends on the Mach number of SNR shocks, $M$. 
According to the standard diffusive shock acceleration (DSA) theory 
\cite{bell78}, the index, $s_i$, is given by 
\begin{eqnarray}
s_i=2 \frac{M^2+1}{M^2-1}~~.
\end{eqnarray}
High energy CRs escape from the CR sources when $M$ is large, and $s_i=2$,
while low energy CRs escape later when $M$ becomes low, and $s_i>2$. 
Then we expect the spectral breaks for all CR species at the same rigidity. 
It should be noted that if one of $s_i, \alpha_i, \beta_i$ and $\gamma_i$ has a rigidity dependence, the steady-state CR spectra have a spectral break (see Section \ref{sec:5}). 

To explain both the helium harding and the concave breaks at $R\sim 300~{\rm GV}$, 
Ref.\cite{Ohira:2010eq} considered a hot gas with temperature of $T=10^6~{\rm K}$ 
and the density profile of helium, 
\begin{eqnarray}
n_{\rm He} \propto r^{-\delta}~~,
\label{eq:delta}
\end{eqnarray}
being different from that of protons, where $r$ is the radius of a system. 
Such plasma is realized in superbubbles (see Sections \ref{sec:inhomo}).
Figure~\ref{fig:spec} shows that our superbubble model 
is in excellent agreement with 
the new data of AMS-02 by using parameters 
$\alpha=6.5, \gamma=0.44, \delta=0.52, \epsilon_{\rm He}=0.4, \epsilon_{\rm C}=0.013$, and $T=10^6~{\rm K}$ (see Ref.\cite{Ohira:2010eq} for detail). 
The similar spectra of helium and carbon suggest
similar density profiles of these elements.
%%%%%%%%%%%%%%%%%%%%%%%%%%%%%%%%%%%%%%%
%%%%%%%%%%%%%%%%%%%%%%%%%%%%%%%%%%%%%%%
\section{Carbon overabundance}
\label{sec:carbon}
There is a standard picture for the carbon overabundance
based on the past observations of CR compositions
\cite{Meyer:1997vz,Ellison:1997an}.
The observational fact is that 
the abundance of refractory elements relative to solar 
is systematically larger than that of volatile ones.
The interpretation of this behavior is that 
the refractory elements condense into dust grains
and the dust grains are injected into the acceleration 
more efficiently than interstellar gas-phase ions
because of their enormous gyroradii.
The sputtering of accelerated dust yield seed ions
and they are accelerated to the observed CR nuclei with overabundance. 
On the other hand, among the volatile elements, 
the abundance enhancements relative to solar increase with mass,
whereas the refractory component shows little mass dependence.
The mass dependence of volatile elements may reflect
the scattering cross section between fast grains and the ambient gas
\cite{Lingenfelter:2007}
or a mass-to-charge dependence
of the injection efficiency via the gyroradius filtering.
Thus the volatile elements are presumably injected directly from the interstellar gas.

CR carbon falls between
the refractory and volatile elements.
Therefore it is reasonable that 
the CR carbon is injected from dust grains as well as gas-phase ions.
This provides an important clue to the origin of CRs 
as discussed in the next section.
%%%%%%%
%%%%%%%
\section{Origin of different spectra}
\label{sec:4}
It is a mystery that the spectral index of protons is different 
from that of helium and carbon
because the same rigidities give the same gyroradii
and hence the usual acceleration and propagation are composition blind.
Since the helium hardening was observed \cite{Ahn:2010gv,Adriani:2011cu,Yoon:2011aa},
several models have been proposed 
for explaining the spectral difference between helium and protons.
These can be broadly divided into four types:
propagation (Section~\ref{sec:4.1}), different sources (Section~\ref{sec:4.2}), injection physics (Section~\ref{sec:4.3}) and inhomogeneous environment (Section~\ref{sec:inhomo}), where the last three are related to the nature of the CR accelerators. 

The carbon spectrum precisely observed by AMS-02 provides a hint to the riddle of the different spectra.
The spectral index of carbon is almost equal to that of helium and strongly suggests that the acceleration site is the same for helium and carbon. 
In addition the amount of injection should keep a constant ratio between helium and carbon even if carbon dust coexists with helium gas (see Section~\ref{sec:carbon}). 
These requirements severely constraint the physical condition of the acceleration site. 
In the following we inspect the models for the spectral differences in the light of the AMS-02 results.
The score sheet for each model is summarized in Table~\ref{tab:score1}. 
%%%%%%
%%%%%%
\subsection{Propagation}
\label{sec:4.1}
The CR compositions do change during propagation through the interactions with the ISM \cite{Blasi:2011fi}. 
However spallation of helium and carbon is not effective under the grammage inferred from B/C and the antiproton flux \cite{Vladimirov:2011rn}. 
Thus, hereafter, we discard this possibility and concentrate on the models related with the CR accelerators. 
%%%%%%
%%%%%%
\subsection{Different sources}
\label{sec:4.2} 
The different spectra may simply reflect different sources, i.e., the proton sources may be naively different from the helium and carbon sources.
For this type of models, in general, the following conditions should be satisfied:
\begin{enumerate}
\item A fine-tuning is necessary in the sense that
completely different sources realize the helium to proton ratio comparable to the solar abundance within a factor of two. 

\item In the helium (proton) sources, the proton (helium) acceleration should be suppressed below the solar abundance. Otherwise the helium to proton ratio would deviate from a power law.

\item The position of the spectral breaks should coincide between protons and helium
even if the sources are different.
This is not trivial since the physical conditions at the acceleration sites are different.
Thus we may have to call for propagation to produce the spectral breaks (see Section~\ref{sec:5})
\end{enumerate}
In the real situation, these conditions seem difficult to realize.
One representative model is the wind model, in which
a supernova explodes into the stellar wind
and the CR helium and carbon are accelerated in the wind,
while the protons are accelerated from ISM \cite{Biermann:2010qn}.
Helium and carbon are abundant in the wind material 
if the central star is the progenitor of type Ib and Ic supernovae, respectively. 
However type II supernovae occur most frequently and the composition of 
their winds is quite similar to that of ISM. 
Thus, the proton acceleration in the wind of type II supernovae should be suppressed 
below the solar abundance by an unknown mechanism. 
Furthermore, in this model the helium acceleration in the ISM should be suppressed 
below solar abundance by another unknown mechanism. 
Therefore, it seems hard to explain both suppressions simultaneously.

The other model is the reverse shock model, in which
the reverse shock running back into the supernova ejecta 
accelerates the helium and carbon \cite{Ptuskin:2012qu}. 
Note that type Ia supernovae do not produce helium and carbon so much.
However the reverse shock is much less energetic than the forward shock
except for a very brief period of the shock-crossing
and accelerated particles may suffer from adiabatic losses.
Moreover, as with the wind model, the helium acceleration should be somehow suppressed below solar abundance in the forward shock by an unknown mechanism.

\subsection{Injection physics}
\label{sec:4.3}
In the injection model, the differences between elements arise from the injection process
in the collisionless shock acceleration \cite{Malkov:2012gb}.
According to the injection model of Ref.\cite{Malkov:2012gb}, the injection into DSA is more efficient at 
smaller Alfv{\'e}n Mach numbers, $M_A$, and protons have the strong dependence on $M_A$ 
compared with helium. 
Besides, the spectra of accelerated particles are softer when the shock is weaker.
Then the time-integrated spectrum of CR protons becomes softer than that of helium, as observed.

However, if this is the case, the carbon to helium ratio deviates from a constant 
because the rigidity of carbon, especially carbon dust (See Section~\ref{sec:carbon}) and partially-ionized carbon ions, is different from that of helium. 
This is inconsistent with the AMS-02 observations. 
Thus the injection physics is not likely the cause of the different spectra.

Note that the rigidity of carbon ions are equal to that of helium if carbons are fully ionized in high temperature,
and the carbon overabundance may be also explained if CRs originate from carbon rich plasma like superbubbles. 
In this case, however, we should change the standard picture for the dichotomy between refractory and volatile elements in Section~\ref{sec:carbon}.
%%%%%%
%%%%%%
\subsection{Inhomogeneous environment}
\label{sec:inhomo}
The energy spectra may differ between elements if the composition of the ambient medium is not uniform 
\cite{Ohira:2010eq,Drury:2010am}. 
Because a stronger shock accelerates higher-energy CRs, the outward-decreasing abundance naturally 
leads to the hard spectrum of CR helium. 
One can easily understand this mechanism in the escape model in which the highest-energy CRs escape from the source and shape the observed energy spectrum. 
Higher-energy helium is produced in the inner region where the helium is more abundant and the shock is stronger. 

The composition gradient may be produced by a gradient of the ionization degree of helium if  neutral helium atoms are not injected into acceleration \cite{Drury:2010am}. 
However, neutral atoms could be injected into acceleration \cite{ohira12}. 
Moreover, the carbon (dust and partially-ionized ions) is expected to have a different ionization profile from that of helium. 
Thereby the carbon to helium ratio is not kept constant. 
This is inconsistent with the AMS-02 observations.

The chemical inhomogeneity is expected in superbubbles. 
A superbubble is created by multiple supernovae and stellar winds in OB associations. 
The supernova ejecta dominates the bubble mass especially in the core region \cite{Higdon:1998}. 
Since the supernova progenitors have too small proper motion to escape the bubble, the supernova shock may sweep the ejecta-enriched matter. 
We emphasize that the ejecta-enriched region is essential for the helium inhomogeneity because the stellar nucleosynthesis can not double the cosmic mean helium abundance because of the Sch\"onberg-Chandrasekhar limit \cite{Ohira:2010eq}. 
This is one of the reason why we need the big bang nucleosynthesis. 
Therefore the superbubble model with concentration of composition is an inevitable consequence of our arguments. 

To reproduce the constant carbon to helium ratio as observed by AMS-02 in Figure~\ref{fig:ratio},
the helium and carbon should be well mixed in the supernova ejecta such as via the Rayleigh-Taylor instability 
or in the superbubble via supernova-driven turbulence. 
Then we also expect similar spectra for heavy metals (possibly except for CR iron which might be produced by type Ia supernovae). 
On the other hand, the proton fraction may be different since the hydrogen is ejected at completely different spacetime and velocities.

More importantly, we can argue that the old ISM with solar abundance should not mix with the fresh supernova ejecta so much in the acceleration region. 
Because primordial helium and carbon mass fractions are $Y_{\rm p}=0.2477\pm0.0029,~Z_{\rm c,p}=0$ \cite{Peimbert07} 
while those fractions in the present ISM are $Y_{\rm ISM}=0.275, Z_{\rm c,ISM}=0.0034245$ \cite{Hirschi05}, 
the mean mass ratio of carbon to helium in stellar ejecta is given by 
\begin{equation}
\frac{Z_{\rm c,ej}}{Y_{\rm ej}}=\frac{Z_{\rm c, ISM}}{Y_{\rm ISM}-Y_{\rm p}} = 0.1254~~,
\end{equation}
which is much larger than that of ISM, $Z_{\rm c,ISM}/Y_{\rm ISM}=0.01245$. 
In order to produce a hard spectrum of CR helium compared to that of CR protons, 
our model needs a helium rich region at the center of superbubbles. 
Therefore, the ejecta should dominate over ISM at the center of superbubbles and 
the carbon to helium ratio, $Z_{\rm c} / Y$, should be close to $Z_{\rm c,ej} / Y_{\rm ej}$. 
If the mass fraction of ISM is larger than that of stellar ejecta in the outer region of superbubbles, 
$Z_{\rm c} / Y$ should be close to $Z_{\rm c,ISM} / Y_{\rm ISM}$ 
and then $Z_{\rm c} / Y$ strongly depends on the superbubble radius. 
If so, the index, $\delta$, in Equation \ref{eq:delta} for helium is different from that for carbon, 
and hence the spectra of escaping CRs differ between helium and carbon. 
However, since the carbon to helium ratio in CRs provided by AMS-02 is constant 
within $\sim 20~\%$ in Figure~\ref{fig:ratio}, 
the mixing fraction of ISM should be less than $\sim 20~\%$ at rigidity $R\sim 10~{\rm GV}$, 
and less than $\sim 10~\%$ at $R\sim 10~{\rm TV}$. 
In other words, CRs originate primarily from the supernova 
ejecta and stellar winds in the interiors of superbubbles.

For the Salpeter initial mass function \cite{salpeter55}, massive stars with about $10~M_{\odot}$ 
mainly contribute to the CR production. 
According to a stellar nucleosynthesis model, stellar ejecta of a non-rotating massive star 
with $12~M_{\odot}$ give 
$Z_{\rm c,ej}/Y_{\rm ej}=0.0298=2.39~(Z_{\rm c,ISM}/Y_{\rm ISM})$ \cite{Hirschi05}, 
which is smaller than the value in the above argument. 
In this case, the mixing fraction of ISM may be about $50~\%$. 
However, we should keep in mind that the theoretical models of the stellar nucleosynthesis have still large uncertainties.

%%%%%%
%%%%%%
\section{Origin of spectral breaks}
\label{sec:5}
Theoretical explanations for the spectral breaks in all the CRs at a single rigidity are classified into three types: the breaks produced intrinsically at the CR sources, the breaks produced during the propagation of CRs in the ISM, and the breaks due to the contribution of a local source at low/high rigidities. 
In the following, we will find that even the current AMS-02 data is not accurate enough to pin down the model for the breaks. 
%%%%
%%%%
\subsection{Intrinsic}
\label{sec:5.1}
There are several mechanisms to produce the breaks at the sources.
As discussed in Section~\ref{sec:2}, the Mach number dependence of the spectral index can produce the breaks within the framework of the standard DSA. 
The reacceleration of pre-existing CRs at the superbubble \cite{Parizot:2004em} or the nonlinear acceleration of CRs \cite{Malkov:2001, Ellison:1997an, Ptuskin:2012qu} are also the scenarios that give spectral breaks at the acceleration sites. 
Even when the CR spectrum at a source has a simple power-law form, if multiple kinds of sources \cite{Zatsepin:2006} or sites \cite{Biermann:2010qn} contribute to the CR flux or each spectral index has a dispersion \cite{Yuan:2011}, the superposed spectra can have breaks. 
B/C would not have a break in these "intrinsic" scenarios, 
although there is an exception if the secondary boron is produced in the source \cite{tomassetti15}. 

%%%%%
%%%%%
\subsection{Propagation}
\label{sec:5.2}
If the energy dependence of the diffusion coefficient has a break at the rigidity $\sim 300~{\rm GV}$, the observed CR spectrum would have a break at the same rigidity \cite{blasi12}. 
In this case, the energy spectrum of B/C would also have a break, and a larger ratio is predicted in the higher energy range \cite{Vladimirov:2011rn}. 
Although such a feature has not been confirmed by AMS-02, we cannot rule out this scenario because of the large uncertainty of data in the high energy range. 
The situation is similar in other propagation models, e.g., a model in which the diffusion coefficient is not separable in energy and space  \cite{Tomassetti:2012}, and in which CRs are reaccelerated before escaping from our Galaxy \cite{thoudam14}. 
%%%%%%
%%%%%%
\subsection{Local source}
\label{sec:5.3}
If there is a local CR source in addition to the Galactic sources, the observed CR spectrum would have a break at a certain rigidity, below/above which the local source dominates the CR flux \cite{Thoudam:2012}. 
The spectral difference between the local source and the Galactic sources may be produced by such as the CR confinement \cite{Kawanaka:2010uj}.
However, \cite{Vladimirov:2011rn} show that the low-energy local source scenario fails to reproduce the observed B/C at low rigidities. 
On the other hand, in the high-energy local source scenario, B/C would be suppressed at high rigidities, where the errorbars of data by AMS-02 are still too large to rule out this scenario. 
In these scenarios, the ratio of CR protons to CR helium (and possibly to CR carbon) produced by the local source should coincide with that by the Galactic sources.
%%%%%%%%
%%%%%%%%
\section{Summary and Discussion}
\label{sec:6}
We discussed that superbubbles are the most plausible source of Galactic CRs to explain the recent results of AMS-02, especially the helium and carbon hardenings. 
The spectra of CR helium and carbon are different from that of CR protons because the abundance profile is not uniform in superbubbles. 
All CR spectra may also break at $R\sim 300~{\rm GV}$ since the shock Mach number evolves in high temperature ($\sim 10^6~{\rm K}$) plasma. 

The superbubble origin of CRs is well motivated by other observations \cite{Montmerle:1979,Higdon:1998}. 
First, superbubbles contain $\sim 75\%$ of all Galactic supernovae \cite{Higdon:2005}. 
Second, the observed abundances of light elements, Li, Be and B, in halo stars require CR carbon and oxygen in the early metal-poor Galaxy and imply the superbubble origin \cite{Ramaty:1996qt}. 
Third, the excess of the isotopic ratio ${}^{22}$Ne/${}^{20}$Ne suggests a link to winds from Wolf-Rayet stars, probably a major source of ${}^{22}$Ne, and thus to superbubbles \cite{Binns:2008}, although this is not supported by the updated wind model \cite{Prantzos:2011rm}. 
Finally, gamma rays are detected from OB associations such as in the Cygnus region \cite{Ackermann:2011} and Westerlund 2 \cite{Aharonian:2007qf}. 
Also theoretically, the particle acceleration is efficient in superbubbles \cite{Bykov:1992,Parizot:2004em}.
CR observations for ${}^{59}$Ni and ${}^{60}$Fe suggest that CRs must be accelerated in 
$10^5~{\rm yr}<t<10^7~{\rm yr}$ after a supernova explosion \cite{wiedenbeck99,israel15}, that is also consistent with the superbubble origin of CRs \cite{Higdon:1998,Binns:2008,israel15}.

We suggested that the supernova ejecta dominate the ISM in the acceleration site to explain the AMS-02 observations of He/$p$ and C/$p$ up to $R\sim1~{\rm TV}$ in Section~\ref{sec:inhomo}. 
On the other hand, some studies based on CR data up to $R\sim 1~{\rm GV}$ imply that the supernova ejecta fraction is $\sim 15$--$25\%$ and the other is mixed from the ISM \cite{Rauch:2009ty,Lingenfelter:2007}, partly because this mixing ratio improves the power-law fit to the relation between the abundance and atomic mass. 
Because these arguments are based on the data in different rigidities, 
the mixing ratio derived in this paper ($R>30~{\rm GV}$) is allowed to be different from that derived in previous studies ($R<1~{\rm GV}$). 
Thus, higher-energy CRs, observed by AMS-02, could be produced in more enriched core of superbubbles.

If the composition of CRs with $R>30~{\rm GV}$ at the source is different from that with $R<1~{\rm GV}$, 
there must be some structures in CR spectra below $R=30~{\rm GV}$. 
However, it is not always observable. 
As discussed in the last paragraph of the section \ref{sec:inhomo}, 
in order to explain the new AMS-02 data, the mixing fraction of supernova ejecta may be about $50~\%$. 
If so, the difference of the fraction between $R<1~{\rm GV}$ and $R>30~{\rm GV}$ 
is only by a factor of 2. 
Interestingly, data of AMS-02 actually suggest that the CR carbon to proton ratio, C/$p$, 
has a signature of deviation from a single power law at $R\sim10~{\rm GV}$ (See Figure~\ref{fig:ratio}). 
This is a kind of signature that supports the change of composition. 
In addition, structures in CR spectra below $R\lesssim 30~{\rm GV}$ could be smeared by the second order acceleration during the Galactic propagation. 
In oder to discuss more quantitatively, we need to calculate the CR propagation 
in our  Galaxy and take into account nonlinear effects, dust acceleration, solar modulation and so on \cite{Ellison:1997an}. 
These will be addressed in future work. 

In the near future, AMS-02 will provide precise data for heavy nuclei, which are predicted 
to have similar spectra in our model (possibly except for CR iron). 
Furthermore, the precise data for heavy nuclei with $R<30~{\rm GV}$ will give us some hints 
for the composition change between $R<1~{\rm GV}$ and $R>30~{\rm GV}$. 
CALET \cite{torii08} and ISS-CREAM \cite{seo14} will also observe
CRs above $R\sim  1~{\rm TV}$. 
The B/C provided by future experiments can establish whether the breaks at $R\sim 300~{\rm GV}$ are due to intrinsic, propagation, or local source origin.

%%%%%%%%%%%%%%%%%%%%%%%
%%%%%%%%%%%%%%%%%%%%%%%
\acknowledgments
We are grateful to the referee for useful suggestions and comments. 
We would like to thank H. Kodama, K. Kohri and R. Yamazaki for useful comments. 
We are also grateful to ISSI (Bern) for support of the team "Physics of the injection 
of particle acceleration at  astrophysical, heliospheric, and laboratory collisionless shocks". 
This work is supported by JSPS KAKENHI Grant Nos.~26287051, 24103006, 24000004 and 26247042 (K.I.). 

%%%%%%%%%%%%%%%%%%
%%%%%%%%%%%%%%%%%%


\begin{thebibliography}{55}

\bibitem{Drury:2012md}
L. O'C. Drury, Astropart. Phys. {\bf 39}, 52 (2012)

\bibitem{Meyer99}
J. P. Meyer and D. C. Ellison, ASP Conference Series {\bf 171}, 187 (1999)

\bibitem{aguilar15} 
M. Aguilar et al. (AMS Collaboration), Phys. Rev. Lett. {\bf 114}, 171103 (2015);
M. Aguilar et al. (AMS Collaboration), Phys. Rev. Lett. {\bf 114}, 211101 (2015);
AMS-02 Collaboration, "AMS Days at CERN'' and Latest Results, 15, April (2015)

\bibitem{Adriani:2011cu} 
O. Adriani et al. (PAMELA Collaboration), Science, {\bf 332}, 69 (2011)

\bibitem{Ahn:2010gv} 
H. S. Ahn et al., Astrophys. J. {\bf 714}, L89 (2010)

\bibitem{Vladimirov:2011rn}
A. E. Vladimirov et al., Astrophys. J. {\bf 752}, 68 (2012)

\bibitem{Lodder:2003zy}
K. Lodders, Astrophys. J. {\bf 591}, 1220 (2003)

\bibitem{Ohira:2010eq}
Y. Ohira and K. Ioka,  Astrophys. J. {\bf 729}, L13 (2011)

\bibitem{Kohri:2015mga}
K. Kohri et al., arXiv:1505.01236 (2015);
Y. Fujita et al., Phys. Rev. D {\bf 80}, 063003 (2009)

\bibitem{ohira10}
Y. Ohira, K. Murase, and R. Yamazaki, Astron. Astrophys. {\bf 513}, A17 (2010)

\bibitem{bell78}
A. R. Bell, Mon. Not. R. Astron. Soc. {\bf 182}, 147 (1978);
R. D. Blandford and J. P. Ostriker, Astrophys. J. {\bf 221}, L29 (1978)

\bibitem{Meyer:1997vz}
J. P. Meyer, L. O'C. Drury, and D. C. Ellison, Astrophys. J. {\bf 487}, 182 (1997)

\bibitem{Ellison:1997an}
D. C. Ellison, L. O'C. Drury, and J. P. Meyer, Astrophys. J. {\bf 487}, 197 (1997)

\bibitem{Lingenfelter:2007}
R. E. Lingenfelter and J. C. Higdon, Astrophys. J. {\bf 660}, 330 (2007)

\bibitem{Yoon:2011aa}
Y. S. Yoon et al., Astrophys. J. {\bf 728}, 122 (2011)

\bibitem{Blasi:2011fi} 
P. Blasi and E. Amato, JCAP, {\bf 01}, 010 (2012)

\bibitem{Biermann:2010qn} 
P. L. Biermann et al., Astrophys. J. {\bf 725}, 184 (2010)

\bibitem{Ptuskin:2012qu}
V. Ptuskin, V. Zirakashvili, and E. S. Seo, Astrophys. J. {\bf 763}, 47 (2013)

\bibitem{Malkov:2012gb}
M. A. Malkov,  P. H. Diamond, and R. Z. Sagdeev, Phys. Rev. Lett. {\bf 108}, 081104 (2012)

\bibitem{Drury:2010am}
L. O'C. Drury, Mon. Not. R. Astron. Soc. {\bf 415}, 1807 (2011)

\bibitem{ohira12}
Y. Ohira, Astrophys. J. {\bf 758}, 979 (2012);
Y. Ohira, Phys. Rev. Lett. {\bf 111}, 245002 (2013);
Y. Ohira, Phys. Astrophys. J. {\bf 817}, 137 (2016); 
Y. Ohira and F. Takahara, Astrophy. J. {\bf 721}, L43 (2010)

\bibitem{Higdon:1998}
J. C. Higdon, R. E. Lingenfelter, and R. Ramaty, Astrophys. J. {\bf 509}, L33 (1998)

\bibitem{Peimbert07} 
M. Peimbert, V. Luridiana, and A. Peimbert, Astrophys. J. {\bf 666}, 636 (2007)

\bibitem{Hirschi05}
R. Hirschi, G. Meynet, and A. Maeder, Astron. Astrophys. {\bf 433}, 1013 (2005)

\bibitem{salpeter55} 
E. E. Salpeter, Astrophys. J. {\bf 121}, 161 (1955)

\bibitem{Parizot:2004em} 
E. Parizot et al., Astron. Astrophys. {\bf 424}, 747 (2004)

\bibitem{Malkov:2001}
M. A. Malkov and L. O' C. Drury, Rep. Prog. Phys. {\bf 64}, 429 (2001)

\bibitem{Zatsepin:2006}
V. I. Zatsepin and N. V. Sokolskaya, Astron. Astrophys. {\bf 458}, 1 (2006)

\bibitem{Yuan:2011}
Q. Yuan, B. Zhang, and X.-J. Bi, Phys. Rev. D {\bf 84}, 043002 (2011)

\bibitem{tomassetti15}
P. Mertsch and S. Sarkar, Phys. Rev. Lett. {\bf 103}, 081104 (2009); 
N. Tomassetti and F. Donato Astrophys. J. {\bf 803}, L15 (2015)

\bibitem{blasi12}
P. Blasi, E. Amato, and P. D. Serpico, Phys. Rev. Lett. {\bf 109}, 061101 (2012)  

\bibitem{Tomassetti:2012}
N. Tomassetti, Astrophys. J. {\bf 752}, L13 (2012)

\bibitem{thoudam14}
S. Thoudam and J. R.  H\"{o}randel, Astron. Astrophys. {\bf 567}, A33 (2014)

\bibitem{Thoudam:2012}
S. Thoudam and J. R.  H\"{o}randel, Mon. Not. R. Astron. Soc. {\bf 421}, 1209 (2012); 
G. Bernard et al., Astron. Astrophys. {\bf 555}, A48 (2013); 
A. D. Erlykin and A. W. Wolfendale, Astropart. Phys., {\bf 35}, 449 (2012)

\bibitem{Kawanaka:2010uj}
N. Kawanaka et al., Astrophys. J. {\bf 729}, 93 (2011)

\bibitem{Montmerle:1979}
T. Montmerle, Astrophys. J. {\bf 231}, 95 (1979)

\bibitem{Higdon:2005}
J. C. Higdon and R. E. Lingenfelter, Astrophys. J. {\bf 628}, 738 (2005)

\bibitem{Ramaty:1996qt}
R. Ramaty, B. Kozlovsky, and R. E. Lingenfelter,  Astrophys. J. {\bf 488}, 730 (1997);
A. Alib\'es, J. Labay, and R. Canal, Astrophys. J. {\bf 571}, 326 (2002)

\bibitem{Binns:2008} 
W. R. Binns et al., New Astronomy Reviews {\bf 52}, 427 (2008)

\bibitem{Prantzos:2011rm} 
N. Prantzos, Astron. Astrophys. {\bf 538}, A80 (2012)

\bibitem{Ackermann:2011}
M. Ackermann M., et al., Science {\bf 334}, 1103 (2011);
A. A. Abdo et al., Astrophys. J. {\bf 58}, L33 (2007)

\bibitem{Aharonian:2007qf} 
F. Aharonian et al. (HESS Collaboration), Astron. Astrophys. {\bf 467}, 1075 (2007)

\bibitem{Bykov:1992}
A. M. Bykov and G. D. Fleishman, Mon. Not. R. Astron. Soc. {\bf 255}, 269 (1992)

\bibitem{wiedenbeck99}
M. E. Wiedenbeck, et al., Astrophys. J. {\bf 523}, L61 (1999)

\bibitem{israel15}
M. H. Istrael et al., Proc. 34th International Cosmic Ray Conference (The Hague), Proceedings of Science, 275 (2015)

\bibitem{Rauch:2009ty}
B. F. Rauch et al., Astrophys. J. {\bf 697}, 2083 (2009); 
{\bf 722}, 970 (2010)[erratum]

\bibitem{torii08}
S. Torii et al. (CALET Collaboration), J. Phys. Conf. Ser. {\bf 120}, 062020 (2008)

\bibitem{seo14} 
E. S. Seo, et al. Adv. Space Res. {\bf 53}, 1451 (2014)

\end{thebibliography}
\end{document}